\documentstyle[aps,epsf,amsmath,amssymb]{revtex}    

\begin{document}

\draft

\title{Density of States of Disordered Two-Dimensional Crystals with
  Half-Filled Band}

\author{E. P.Nakhmedov $^{1,2)}$ , M.Kumru $^{3)}$ and R.Oppermann $^{1)}$ }
\address{1) Institut f\"ur
    Theoretische Physik, Univ. W\"urzburg,D--97074 W\"urzburg, FRG \\
  2) Azerbaijan Academy of Sciences,Institute of Physics,H.Cavid st.33,Baku,Azerbaijan\\
3)Fatih  University, B\"uy\"ukcekmece, Istanbul,T\"urkiye}

  \date{9 July 1999}

    \maketitle

\begin{abstract}

A diagrammatic method is applied to study the effects of
commensurability in two-dimensional disordered crystalline metals by
using the particle-hole symmetry with respect to the $ nesting $  vector
${\boldsymbol P_0}={\{\pm \frac {\pi}{a}, \frac {\pi}{a} \} }$ for a half-filled electronic band. The density of electronic states (DoS) is shown to have  nontrivial quantum corrections due to both $ nesting $ and elastic impurity scattering processes, as a result the van Hove singularity is preserved in the center of the band. However, the energy dependence of the DoS is strongly changed. A small offset from the middle of the band gives rise to disappearence of quantum corrections to the DoS .

\end{abstract}


\narrowtext Great advance has been made in the theory of
disordered metals after the pioneering  work of Abrahams et
al.\cite{abrahams79}. According to this paper all electronic
states in one- and two-dimensional (1d and 2d) disordered systems are
localized irrespective of the degree of randomness,\cite{mott61a,berezinskii73a}. Electron-electron correlations in disordered metals have
been shown to result in nontrivial corrections to the density of
electronic states,(DoS), and conductivity,\cite{altshuler79,altshuler80,altshuler85,fukuyama85,lee85}. The
corrections to the conductivity are similar to the localization
corrections obtained for a noninteracting electron gas, and quantum
corrections to the DoS were shown to reduce it near the Fermi level.

Disordered metals in all abovementioned papers are modeled as a free
electron gas moving in the random field of rigid impurities. However,
at low concentrations of impurities the crystal usually has a
periodical structure and the impurity atoms in most cases substitute the host atoms of the lattice. Then, 
the effects of commensurability of the electron wavelength, $\lambda$ , and the lattice constant, $a$, become essential in the scattering processes. The commensurability is known to exist at all 'rational' points of the electron band,however it appears to be important for a half-filled band.

In this Letter we present our study concerning  the  effects of weak disorder on the electronic DoS of a 2d crystalline metal.The Hamiltonian of the model can be written as $\hat{H}= \hat{H_o} + V({\bf r})$, where $\hat{H_o}$ is the Hamiltonian of noninteracting electrons in the perfect square lattice with nearest- neighbor hopping and $ V({\bf r}) = \sum_{i} U({\bf r} - {\bf R_{i}})$ is the impurity potential with ${\bf R_{i}}$ being the positional vector of an impurity randomly located on the $i-th$ lattice site. 

The one-particle DoS of the regular lattice can be expressed as $\rho_0^{(d)}=\frac{2}{(2\pi\hslash)^d}\int\frac{d\boldsymbol S}{\arrowvert\nabla\epsilon(\boldsymbol k)\arrowvert}$ , where $ d{\boldsymbol S}$ is the element of an isoenergetical surface in d-dimensional space. It can be shown that $ \rho_0^{(d)}(\epsilon)$  has a van Hove singularity at the points where the group velocity of the electron wave packet ${\boldsymbol{\it V}_{\boldsymbol k}}= \nabla \epsilon ({\boldsymbol k})$ vanishes,\cite{ashcroft76}. For a  three-dimensional 
 (3d) regular lattice $\rho_0^{(3)}(\epsilon)$ has integrable singularities.

For a pure 2d lattice with nearest-neighbor hopping the van Hove singularity has logarithmic character.
The simplest electron spectrum for a 2d square lattice can be written in the tight-binding approximation as 
\begin{equation}
\epsilon (\boldsymbol k) = t [ 2 - \cos (k_xa) - \cos (k_ya)];
\end{equation} 

where,$k_{x,y}=\frac{2\pi}{aN_{x,y}}n_{x,y}$ with $-\frac{N_{x,y}}{2}< n_{x,y}\leq \frac {N_{x,y}}{2}$ and only electron tunneling between  nearest-neighboring sites with the tunneling integral $t$ is involved. The bandwidth is $ W=4t$; and for a half-filled band case the Fermi energy becomes $\epsilon _F =2t$. The DoS of a 2d square lattice with nearest-neighbor hopping is expressed by the elliptic integral of the first kind, an asymptotic expression of which has a logarithmic singularity in the middle of the energy band as:
\begin{equation}
\rho_0^{(2)}(\epsilon)=
     \begin{cases}
\frac{1}{{(\pi a)}^2\sqrt{\epsilon_F^2-{\arrowvert \tilde{\epsilon} \arrowvert}^2}} \ln \lgroup \frac {4t^2-|\tilde{\epsilon }|^2}{|\tilde{\epsilon }|^2} \rgroup & \arrowvert \tilde{\epsilon} \arrowvert \neq 2t,\\
\frac{2}{{(\pi a)}^2 \arrowvert \tilde{\epsilon} \arrowvert },& \tilde \epsilon \Rightarrow \pm 2t
\end{cases}
\end{equation}
where $\tilde{\epsilon}$ is an electron energy measured from the Fermi level $\tilde{\epsilon}=\epsilon - 2t$.(Hereafter the tilde on $\epsilon $ will be dropped).

The DoS of a noninteracting electron gas moving in the random field of impurities has no essential singularities near the Fermi surface. Inclusion of even short-range correlations in the 2d disordered metal gives rise to a decreasing DoS near the Fermi level,\cite{altshuler80,altshuler85,finkelshtein83}.       
 The DoS of a $2d$ disordered crystal with substitutional impurities turns out to have a singularity near the middle of the band even for the noninteracting electron gas.

Notice that the commensurability effect for a $1d$ disordered crystal near the middle of the band has been studied by many authors,\cite{dyson53,weissman75,gor'kov76,ovchinnikov77,gogolin77,gredeskul78,hirsch76,eggarter78,kozlov98}. Dyson first pointed out \cite{dyson53} that the phonons' DoS of a $1d$ disordered chain has a singularity as $\rho^{(1)}(\epsilon ) \propto - {\arrowvert \epsilon \arrowvert}^{-1} \ln ^{-3} \arrowvert \epsilon \arrowvert $ near the middle of the band. Later an analogous singularity has been found in the electronic DoS of many $1d$ models,\cite{weissman75,gor'kov76,ovchinnikov77,gogolin77,gredeskul78,hirsch76,eggarter78}. However, there exist a few computational studies of the DoS of a $2d$ disordered crystal,\cite{tsujino79,soukoulis82,hu84,eilmes98}. By studying the averaged Green's function for a disordered system with $n$ orbitals per site,the expansion coefficients in power of $1/n$ for $d \geq 2$ were shown in \cite{oppermann79a} to diverge for energies approaching  the band center. This fact was interpreted in \cite{oppermann79a} as a existence of a van Hove singularity in the DoS.

The technical difficulties in analytical calculations of physical parameters are connected with the cosine energy spectrum and, furthermore, with an absence of perturbative parameters for half-filling.
It is appropriate to notice that the recent attempt to explain both the linear resistivity and the nearly temperature independence of the thermopower in the cuprate superconductors \cite{newns94} is based on the existence of a van Hove singularity in the DoS of these materials. To study the problem analytically, the idealized model \cite{dzyaloshinskii87,pattnaik92} $\epsilon (\boldsymbol k)= k_xk_y$ for the energy spectrum, which gives a logarithmic DoS at half-filling, was used. There exists also the $nested$ Fermi Liquid  scattering approach \cite{virosztek90,ruvalds95} to study the susceptibility of high-$T_c$ superconductors.
 
As it is known, the Fermi surface of an infinite $2d$ lattice with nearest-neighbor hopping is changed with band-filling and it is flat for the half-filling. In this case there exists a $nesting$ vector $\boldsymbol P_o=\{\pm \frac {\pi}{a}, \frac {\pi}{a}\}$ that maps an entire section of the Fermi surface onto another, i.e. the Fermi surface is perfectly  $ nested $  for the half-filled band case. There, the following particle-hole symmetry of the electron dispersion with respect to the vector $\boldsymbol P_o $ for a half-filled band holds:
\begin {equation}
\epsilon (\boldsymbol p + \boldsymbol P_o)  - \epsilon_F = - [\epsilon (\boldsymbol p) - \epsilon _F]
 \end{equation}

The one-electron DoS  can be calculated according to the following expression:
\begin{equation}
\rho(\epsilon ) = - \frac{2}{\pi}{\rm Im}\int \frac {d^2 p}{(2\pi)^2} G_R(\boldsymbol p,\epsilon)
\end{equation}
where $G_R(\boldsymbol p,\epsilon )$ is the retarded Green's function.

\begin{figure}
\epsfxsize8.5cm
\centerline{\epsfbox{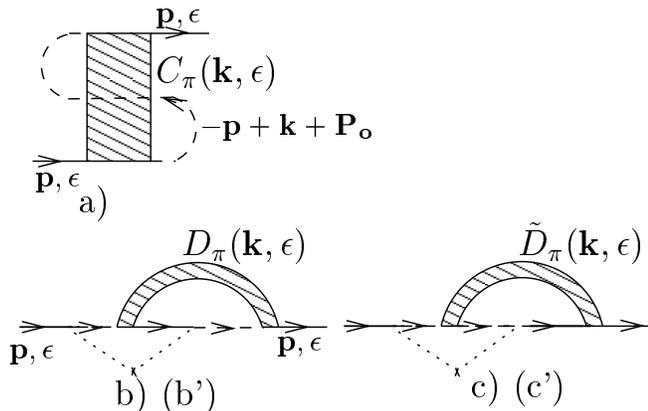}}
\caption{Self-energy parts which give a contribution to the Green's function. Impurity scattering is denoted by a dotted line with a cross. $C_{\pi}$, $\tilde{C}_{\pi}$ and $D_{\pi}$, $\tilde{D}_{\pi}$ are cooperon and diffusion blocks, respectively. The diagrams $b')$ and $c')$  are the symmetrical conjugate to those shown by $b)$ and $ c)$  with respect to the additional impurity line.}
\label{fig:nesting}
\end{figure}

The new class of diagrams which gives an essential contribution to the DoS is drawn in Fig.\ref{fig:nesting}. Thin solid  and dashed lines in Fig.1 correspond to the 'bare' retarded Green's functions  ${G_R}^o(\boldsymbol p, \epsilon)$ and ${G_R}^o(\boldsymbol p + \boldsymbol P_o, \epsilon )$ .

Averaging over impurity realization is performed according to the crossed diagram technique in the Born approximation described in \cite{Abrikosov75a}, i.e. the white-noise impurity potential is used which corresponds to the off- diagonal disorder in the problem. The main diagrams are selected according to the condition of $\epsilon_{F} \tau \gg 1$ or $k_{F}l \gg 1$ with $l=v_{F}\tau$ and $\tau$ being an elastic scattering time, which means $l\gg \lambda$.
  
The scattering processes in the problem are described by two relaxation times $\tau _o$ and $\tau _{\pi}$, (Fig.2). $\tau _o$ corresponds to the  normal scattering process  contribution to which gives the  diagram shown in Fig.2a  and $ \frac {1}{\tau _o} = \frac{C_{imp}}{(2\pi)^2} \int \frac {d \boldsymbol S}{\arrowvert \boldsymbol {V_k} \arrowvert} {\arrowvert U(S) \arrowvert}^2 $,\cite{Abrikosov75a}. Here, $C_{imp}$ and $U$ are the impurity concentration and potential, respectively. The diagrams which give a contribution to Green's functions contain new impurity vertices, shown in Figs.2b,c. These vertices characterize simultaneous reflections of electrons on the Brillouin zone boundary in the process of scattering on an impurity. The following expression represents these vertices: $\frac {1}{\tau_{\pi}} = \frac {C_{imp}}{(2\pi)^2} \int \frac {d \bf S}{ | \bf {V_k} |}{|U(P_o,S)|}^2$. It can be seen from Figs.2b,c that the total momentum, generally speaking, is not conserved for these impurity vertices. Therefore they represent an $Umklapp$ process.

\begin{figure}
\epsfxsize8.5cm
\epsfbox{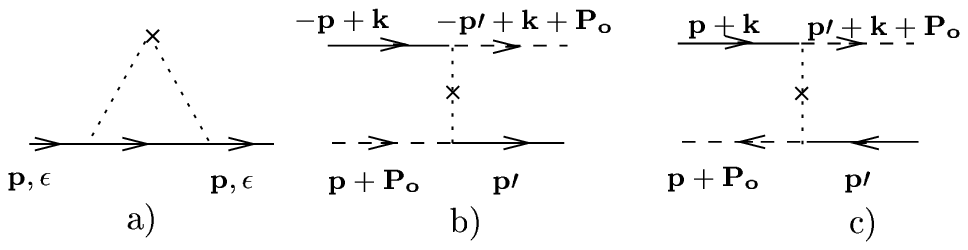}
\caption{}
\label{fig:nesting2}
\end{figure}

Bare retarded and advanced Green's functions  $G_{R,A}^o(\boldsymbol p,\epsilon) = [\epsilon - (\epsilon(\boldsymbol p) - \epsilon_F) +\frac{i}{2\tau} {\rm sign} \epsilon ]^{-1} $ contain a total relaxation time, which is equal to $\frac{1}{\tau}=\frac{1}{\tau _o} + \frac{1}{\tau _{\pi}} $.

The maximally crossed diagrams are redrawn in Fig.1a so that the self-energy part contains the particle-particle propagator $C_{\pi}(\boldsymbol k,\epsilon )$which we  call $\pi$-cooperon, (Fig.3a). In contrast to the cooperon for an isotropic system  $C_{\pi}(\boldsymbol k,\epsilon )$ has a diffusion pole at a small total energy and a large total momentum $\sim {\bf P_o}$. In the limit of small $\epsilon$ and $k$ , satisfying the conditions   $|\epsilon |\tau \ll 1$ and $ kl \ll 1$, $C_{\pi}(\boldsymbol k,\epsilon )$ has the following form
\begin{equation}
C_{\pi}(\boldsymbol k,\epsilon)=\frac{C_{imp}|U(P_o)|^2 }{1-4i\tau |\epsilon | +(kl)^2-(\tau / \tau_{\pi})^2} ;
\end{equation}

which has a diffusion pole for $\tau_{\pi} \to \tau $. Notice that $\epsilon$ is an energy measured from the Fermi level.

The self-energy parts in Figs.1 $b (b')$ and $ c (c')$  contain a particle- hole propagator $D_{\pi}(\boldsymbol k,\epsilon )$ which can be written for $|\epsilon |\tau \ll 1$ and $kl \ll 1$  as
\begin{equation}
D_{\pi}(\boldsymbol k,\epsilon )=\frac{C_{imp}|U(P_o)|^2}{1-4i\tau |\epsilon |+(kl)^2- (\tau / \tau_{\pi})^2} 
\end{equation}

$\pi$- diffuson $D_U(\boldsymbol k,\epsilon) $  has a diffusion pole in the particle- hole channel at a small total energy and large momenta difference $\sim {\bf P_o}$ for $\tau \rightarrow \tau _{\pi}$.

According to the Dyson equation,  the retarded Green's function $G_R(\boldsymbol p,\epsilon )$ is expressed in the following form:          
\begin{equation}
G_R(\boldsymbol p,\epsilon )= \frac {1}{(G_R^o(\boldsymbol p,\epsilon ))^{-1} -\Sigma(\boldsymbol p,\epsilon )}= \sum _{n=0}^{\infty }(G_R^o(\boldsymbol p,\epsilon ))^{n+1}(\Sigma(\boldsymbol p,\epsilon ))^n
\end{equation}
where, $\Sigma(\boldsymbol p,\epsilon )= \sum _{i=a}^{c^{\prime }} \Sigma _i(\boldsymbol p,\epsilon )$, and each of $\Sigma_i(\boldsymbol p,\epsilon )$ corresponds to the one of self-energy parts $a-c^{\prime }$ in Fig.1, respectively. By summing all diagrams in Fig.1 the self-energy part $\Sigma ({\bf p},\epsilon )$ is reduced to the form:
\begin{equation}
\begin{split}
\Sigma ({\bf p},\epsilon ) = & \int \frac {d^2k}{(2\pi)^2} \{ C_{\pi}(\boldsymbol k,\epsilon ) - \frac {2\tau }{\tau _{\pi}}(1 - \frac {\tau }{\tau _{\pi}})D_{\pi}(\boldsymbol k,\epsilon ) \}\\
& {G^o}_R (-{\bf p}+{\bf k}+{\bf P_o},\epsilon);
\end{split}
\end{equation}

\begin{figure}
\epsfxsize8.5cm
\epsfbox{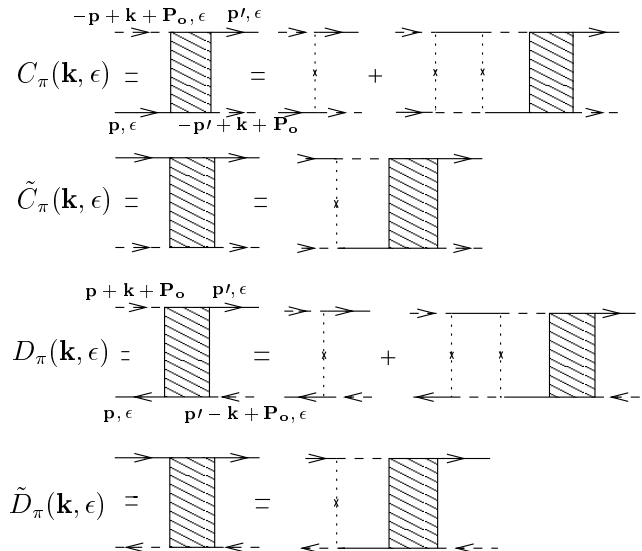}
\caption{Ladder diagrams for the $\pi$- cooperons 
$C_{\pi}({\bf k},\epsilon )$  and $\tilde{C}_{\pi}({\bf k},\epsilon )$, and the $\pi$- diffusons $D_{\pi}({\bf k},\epsilon )$ and $\tilde{D}_{\pi}({\bf k},\epsilon )$. $\tilde{C}_{\pi}({\bf }k,\epsilon )$ and $\tilde{D}_{\pi}({\bf k},\epsilon )$ are expressed by $C_{\pi}({\bf k},\epsilon )$ and $D_{\pi}({\bf k},\epsilon )$ as $\tilde{C}_{\pi}({\bf k},\epsilon )= - \frac{\tau }{\tau _{\pi}} C_{\pi}({\bf k},\epsilon )$ and $\tilde{D}_{\pi}({\bf k},\epsilon ) = - \frac{\tau}{\tau _{\pi}} D_{\pi}({\bf k},\epsilon )$ for small ${\bf k} $ and $\epsilon $.}
\label{fig:nesting3}
\end{figure}

Substituting Eqs.(7) and (8) into Eq.(4) the following expression for $\rho (\epsilon )$ can be obtained after integrating  the bare Green's functions over ${\bf p}$ and taking a sum over $n$ :
\begin{equation}
\rho (\epsilon )=\rho _o^{(2)}\{ 1- \rm Re \frac {4{\tau }^2 \alpha (\epsilon )}{\sqrt {1+4{\tau }^2 \alpha (\epsilon )}(1+\sqrt {1+4{\tau }^2 \alpha (\epsilon )} ) } \}
\end{equation}
where
\begin{equation}
\alpha (\epsilon )= \int \frac {d^2k}{(2\pi )^2}\{ C_{\pi}(\boldsymbol k,\epsilon )-\frac {2 \tau }{\tau _{\pi}}(1 - \frac {\tau }{\tau _{\pi}})D_{\pi}(\boldsymbol k,\epsilon ) \} ;
\end{equation}

The obtained expression for the DoS contains high order  logarithmically divergent contributions. So, infinite order impurity blocks are summed up according to Eqs.(7)-(8).

The  $nesting$ processes  are weakened away from the middle of the band. In this case the normal scattering processes become  more probable and $\tau _o \ll \tau _{\pi}$ . As a result the diffusion poles in the propagators $C_{\pi}(\boldsymbol k,\epsilon )$ and $D_{\pi}(\boldsymbol k,\epsilon )$ disappear. Therefore a small offset from the middle of the band gives rise to the disappearence of the quantum corrections to the DoS and $\rho (\epsilon )=\rho _o^{(2)} $.
When filling reaches the center of the band the Fermi surface becomes flat. In this case the scattering with $nesting$ seems to have  preference, i.e. $\tau _{\pi} \ll \tau _o $ and $\tau =\tau _{\pi}$. The impurity blocks $C_{\pi}(\boldsymbol k,\epsilon )$ and $D_{\pi}(\boldsymbol k,\epsilon )$  have a diffusion pole under this condition and we get from Eq.(10):
\begin{equation}
\alpha (\epsilon )=\frac {1}{8\pi \epsilon _F {\tau _{\pi}}^3} \ln \lgroup {-\frac {1}{4i\tau _{\pi} |\epsilon |}} \rgroup
\end{equation}

Substitution this expression for $\alpha (\epsilon )$ into Eq.(9) gives the following expression for the DoS:
\begin{equation}
\rho (\epsilon )= \rho _o^{(2)}(\epsilon )\{1-\frac {\alpha_o (\epsilon )}{\sqrt {1+\alpha_o (\epsilon )} [1+\sqrt {1+\alpha_o (\epsilon ) }] } \};
\end{equation}
 
where,$\alpha_o(\epsilon )= \frac {1}{2\pi \epsilon _F \tau _{\pi}} \ln \lgroup \frac {1}{4\tau _{\pi} |\epsilon |} \rgroup $. Near the vicinity of the Fermi level Eq.(12) is approximated as
\begin{equation}
\rho (\epsilon )= \rho _o^{(2)} (\epsilon ) [\frac {1}{2\pi \epsilon _F \tau _{\pi}} \ln \lgroup \frac {1}{4\tau _{\pi} |\epsilon |} \rgroup ]^{-1/2}
\end{equation}

$\rho _o^{(2)}(\epsilon )$ in Eqs.(12) and (13) is the DoS of a pure 2d square crystal, which has the van Hove singularity expressed by Eq.(2). Elastic  scatterings in the crystal with substitutional impurities preserve the central peak of the DoS, however the energy dependence of $\rho (\epsilon )$ is changed from logarithmic dependence to the square root of the logarithm in the close vicinity of the middle of the band:
\begin{equation}
\rho (\epsilon ) \Rightarrow \frac{2}{{(\pi a)}^2 \epsilon_F }{ \lgroup 2\pi \epsilon _F \tau _{\pi} \rgroup}^{1/2} \ln ^{1/2} \lgroup \frac {1}{4\tau _{\pi} | \epsilon |} \rgroup      \quad\mbox{as}\quad |\epsilon |\rightarrow 0
\end{equation}
 
The peak also becomes  narrower than that of the van Hove one due to impurity scattering.

 It is worth to compare here the result presented in the Letter with that obtained for a 1d lattice with half filled energy band, containing off-diagonal disorder,\cite{gogolin77,weissman75,gor'kov76,ovchinnikov77,gredeskul78,hirsch76,eggarter78}. The DoS of the pure 1d lattice is a smooth function of the energy within the band and it has a singularity only at the boundary of the energy band. Therefore the Dyson peak of the DoS in the middle of the half-filled band of the 1d disordered lattice  is a result of strong Bragg reflection of the electrons in the process of scattering on impurities with a consequent interference of scattered waves. The calculation of density-density and current-current correlators for 1d disordered chains shows that the localization length diverges  and static conductivity saturates to a constant value at $T=0$ with the approaching of half filling. In contrast to the 1d case the singular enhancement of the DoS in the 2d disordered lattice is due to a van Hove singularity in the bare DoS. Effect of impurity tends to decrease the DoS. The obtained result is also confirmed by the computational study  of the DoS of a 2d square lattice with off-diagonal disorder \cite{eilmes98}. Numerical analysis of the participation numbers in \cite{eilmes98} shows that localization becoms stronger close to the band center.

Notice that a nested Fermi surface with a nesting vector ${\bf P}^*= \xi {\bf P_o}$ ($\xi<1$) \cite{ruvalds95} arise for a tight-binding models which include both nearest neighbor and next nearest neighbor hopping terms and correspond to a non-half filled band. In this case the van Hove singularity in the DoS of pure system disappears and the DoS vanishes on the Fermi surface.
  
In conclusion, we proposed a new diagrammatic approach to study $nesting$ effects on the DoS of 2d disordered square lattice. Calculations were performed in the diffusion approximation. However, unlike the ``conventional'' localization theory, the infinite order logarithmically divergent diagrams were summed up here according to the Dyson equation. The quantum corrections tend to decrease the DoS in the whole energy interval. As a result, the van Hove singularity  in the middle of the band is preserved, however, its energy dependence is changed from logarithmic to square root of logarithm in the close vicinity of the band center. The calculations are valid for a half-filled energy band. A small shift from the band center results in the vanishing of the quantum corrections to the DoS.

The authors are indebted to a referee who pointed out the difference between impurity effects in 1d and 2d systems. One of the authors (E.P.N.) thanks A.Erzan and all members of the Physical Department of Istanbul Technical University, where the part of this work was performed, for long time hospitality.
E.P.N. also thanks H.Feldmann for discussion.




\end{document}